# Reinventing the Pocket Microscope

*Tahseen Kamal, Xuefei He & Woei Ming Lee*

**The key to the success of pocket microscopes stems from the convenience for anyone to magnify the fine details (tens of micrometres) of any object on-the-spot. The capability with a portable microscope lets us surpass our limited vision and is commonly used in many areas of science, industry, education. The growth of imaging and computing power in smartphones is creating the possibility of converting your smartphone into a high power pocket microscope. In this article, we briefly describe the history of pocket microscopy and elucidate how mobile technologies are set to become the next platform for pocket microscopes.**

## Pocket Microscope

Charles Darwin and Robert Brown[1] undoubtedly benefited from unsophisticated mobile microscopes that gave them the necessary tools to analyse the fine details in their specimens. The widespread popularity of the drum microscope (as depicted in Figure 1) in the mid-18th Century is credited to its portability and ease of use, hence earning the name - the "pocket microscope"[2].

Owing to their small size, pocket microscopes are widely used in the field, this reached its height during the Victorian era of scientific exploration in botany, biology etc. In the later part of 1800s, Abbe[3] found the limits of resolving power of lenses and also led the design of more powerful lens designs which are used today in modern microscopes. These lenses are engineered to operate close to the optical diffraction limits. Now, optical microscopes have grown both in size and price this has made them less accessible to the general public. Modern pocket microscopes are typically equipped with inexpensive moulded polymer lenses instead of hand-polished soda glass, artificial lights from light emitting diodes (LEDs) and a digital camera unit. These microscopes are predominantly used in non-scientific professions ranging from tradesmen, jewellers to school teachers and hobbyists. We can credit the popular appeal and continuing commercial success of pocket microscopes to the simple curiosity of peering into the microscopic world anywhere and anytime.

## The age of smartphone

Modern smartphone devices possess over 3 orders of magnitude higher computing power than the Apollo moon landing computers[4]. The sheer amount of computing power has paved way to a new breed of software, *Apps* provide an easy platform to share, communicate and interact on a global scale anywhere. Meanwhile, the accompanying optical devices, such as charge coupled devices

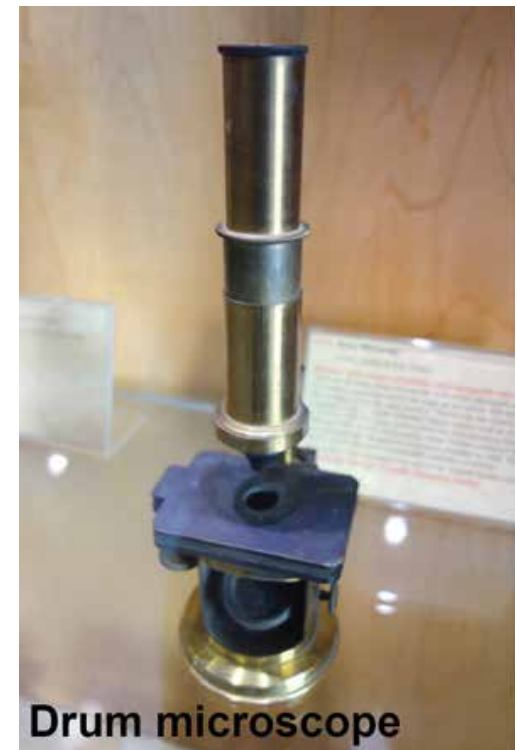

**Figure 1:** *1843 circa, drum microscope, Nachet & Son, France - Golub Collection of Antique Microscopes (Source: http://commons.wikimedia.org/)*

(CCD) camera and LEDs, are getting smaller and better. A smartphone camera nowadays comes equipped with densely packed photo-detectors (~ 2 micrometre per pixels) covering an area of less than 25 $mm^2$ as well as a miniature LED with high brightness (3200mcd). The amount of sensors (light, magnetic, motion, pressure, global positioning unit) on a smartphone combined with the right apps and testing assays enables anyone to own a mobile laboratory[5,6]. It is no coincidence that the demands of these technologies are driving development towards higher quality miniature optics that are approaching close to the optical performances of microscope lenses[7]. In the modern age of technological and information, the dissemination of knowledge through the internet is breaking down traditional ways of learning. Hence, there is a compelling reason to make science more accessible to public and that includes access to microscopes.

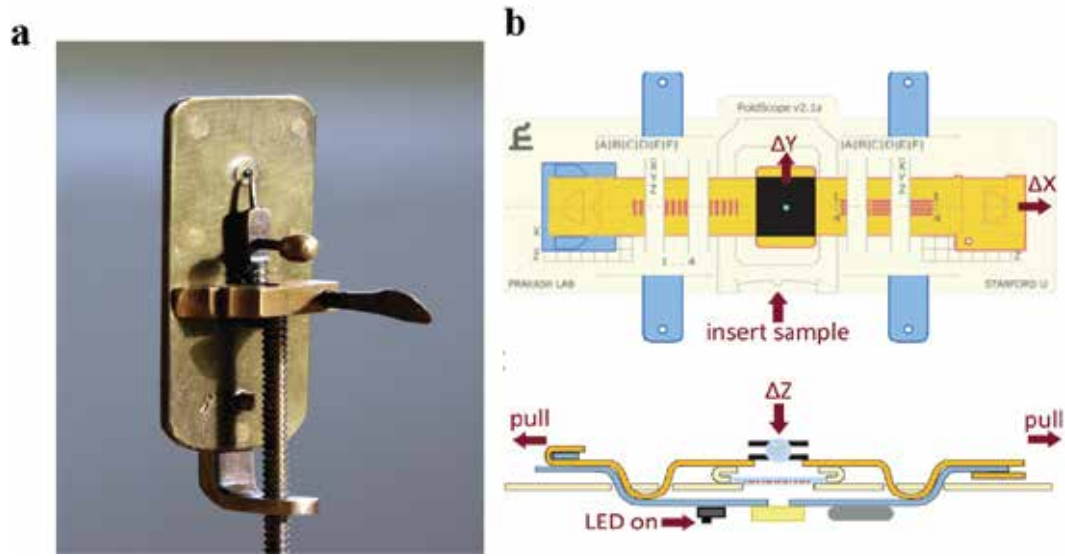

**Figure 2:** *Handheld microscopes (a) Leeuwenhoek's microscope[10] and (b) Foldscope by Cybulski et al[7]*

A multi-purpose smartphone based microscope, coupled with the right apps, can capture samples (biological, material etc.) or be used in diagnosing diseases and even as an educational tool to inspire the next generation of scientists. Hence, optical scientists have begun to explore various ways to shrink the microscope back into our pockets using smartphone technologies[6].

### Lens versus Lensless

Lenses form the most important elements of almost all standard microscope systems. Lenses redirect light that reflects off/transmits through an object onto a secondary plane where a virtual copy of the object is reconstructed. In microscopy, just like a simple magnifying glass, a short focal length imaging lens creates a larger copy (image) of the object that is relayed onto our eyes or a digital camera. In refractive lenses, the focal length is intricately linked to the curvature of the lenses. Leeuwenhoek empirically demonstrated that the imaging power of single well-polished highly curved glass lenses reached down to the resolution of a few micrometres (human red blood cells) and could be mounted onto a home-made hand-held microscope (Figure 2(a)). Reminiscent of Leeuwenhoek's technique, Manu Prakash and his team from Stanford University used commercial glass lenses and miniature LEDs mounted on a paper-frame (paper microscope)[7] to illuminate microscopic objects and project the images either to the eye or a screen for shared viewing (Figure 2(b)).

Since smartphone cameras now have sufficient pixel resolution for high quality imaging, the use of short focal length miniature lenses (glass and plastic) as auxiliary microscope objectives to the imaging lens seems straightforward. More recently, home-made elastomer lenses[8] have also been used to create microscope lenses for smartphones (Figure 3). Lenses can instantaneously relay images, but they possess many optical imperfections (geometric and chromatic aberrations) that need to be corrected with multiple lens elements, which are costly. Lately, there has been an emergence of computational optics (lensless imaging) based on the principles of holography. In computational optics, only a series of blurry images of the object are required. This is because each image (slightly blurred) is taken with light rays illuminating the sample at a slightly different position. With sufficient knowledge of the angle of each illuminating light rays, it is then possible to reconstruct a high resolution image with the series of blurred images. This form of lensless imaging computational technique can be easily implemented on smartphones as shown recently by Lee and Yang[9] who developed a smartphone based chip-scale microscope using ambient illumination which requires no external light-source. By removing the smartphone lens and then tilting the smartphone, they capture multiple separate low resolution images and that is then collate them to produce a high resolution image using super-pixel routines. This technology can greatly simplify the hardware design and make microscopy more accessible to the masses.

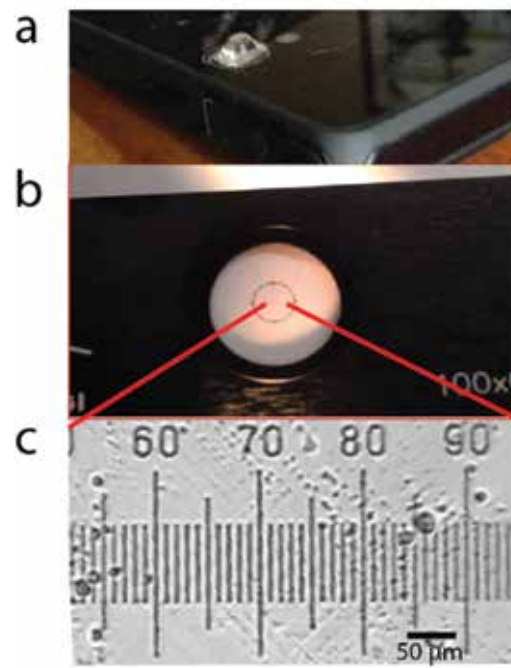

**Figure 3:** *Droplet lens imaging microscope graticule. a) A droplet lens place in front of a smartphone camera. b) Digital photo of microscope graticule (10 μm/division) slide taken with smartphone camera. c) Micrograph of microscope graticule taken with droplets lens attached to the smartphone (transmitted light).*

### Smartphone Pocket Microscope

Digital technology is already transforming the field of pathology where flatbed scanners are used to capture high definition images of excised tissues. The pervasiveness of smartphones (one in every five people owns a smartphone) paves the way for any tablet/smartphone based microscopy system to be transformed into a high resolution digital microscope.

**Woei Ming Lee** (steve.lee@anu.edu.au)
College of Engineering and Computer Science,
Australia National University